\documentclass{paper}
\pdfoutput=1
\usepackage[utf8]{inputenc}
\usepackage[margin=2.5cm]{geometry}
\usepackage{graphicx}
\usepackage[font+=footnotesize]{subcaption}
\usepackage[font=footnotesize]{caption}
\usepackage{url}
\usepackage{multirow}
\usepackage{hyperref}

\title{The Preliminary Evaluation of a Hypervisor-based Virtualization Mechanism for Intel Optane DC Persistent Memory Module}
\author{Takahiro Hirofuchi and Ryousei Takano\\
Information Technology Research Institute,\\
National Institute of Advanced Industrial Science and Technology (AIST)\\
E-mail: t.hirofuchi@aist.go.jp}

\hypersetup{pdfauthor={Takahiro Hirofuchi and Ryousei Takano},pdftitle={The Preliminary Evaluation of a Hypervisor-based Virtualization Mechanism for Intel Optane DC Persistent Memory Module}}

\begin{document}
\maketitle

\begin{abstract}
Non-volatile memory (NVM) technologies, being accessible in the same manner as DRAM,
are considered indispensable for expanding main memory capacities.
Intel Optane DCPMM is a long-awaited product that
drastically increases main memory capacities. However,
a substantial performance gap exists between DRAM and DCPMM.
In our experiments,
the read/write latencies of DCPMM were 400\% and 407\% higher than those of DRAM, respectively.
The read/write bandwidths were 37\% and 8\% of those of DRAM.
This performance gap in main memory presents a new challenge to researchers;
we need a new system software technology supporting emerging hybrid memory architecture.
In this paper, we present
RAMinate, a hypervisor-based virtualization mechanism for hybrid memory systems, and a key technology to
address the performance gap in main memory systems.
It provides great flexibility in memory management and maximizes the performance of virtual machines (VMs) by dynamically optimizing memory mappings.
Through experiments, we confirmed that
even though a VM has only 1\% of DRAM in its RAM, the performance degradation
of the VM was drastically alleviated by memory mapping optimization.
The elapsed time to finish the build of Linux Kernel in the VM was 557 seconds,
which was only 13\% increase from the 100\% DRAM case (i.e., 495 seconds).
When the optimization mechanism was disabled, the elapsed time increased to 624
seconds (i.e. 26\% increase from the 100\% DRAM case).
\end{abstract}


\section{Introduction}
Data centers using virtualization technologies obtain great benefit from increased main memory capacities.
Larger memory capacities of physical machines (PMs) will allow IaaS service providers to consolidate more virtual machines (VMs) onto a single PM,
increasing the competitiveness of their commercial services.
Since DRAM technology is unlikely able to meet this growing memory demand,
non-volatile memory (NVM) technologies, being accessible in the same manner as DRAM,
are considered indispensable for expanding main memory capacities.

RAMinate~\cite{raminateurl, socc2016raminate} is our hypervisor-based virtualization mechanism for the use of such
byte-addressable NVM device in the main memory of a computer.
It creates a unified RAM for a VM from DRAM and NVM,
and dynamically optimizes its memory mappings so as to place hot memory pages in the faster memory device (i.e., DRAM in the current system).
It provides great flexibility in the management of hybrid memory systems.
To the best of our knowledge, RAMinate is the first virtualization mechanism implemented in the hypervisor layer.

In April 2019, Intel officially released the first commercially-available,
byte-addressable NVM technology, Intel Optane Data Center Persistent
Memory Module (DCPMM). DCPMM is a long-awaited product drastically increasing main memory capacities.
However, there is a substantial performance gap between DRAM and DCPMM, which
necessitates novel system software technologies maximizing performance of hybrid memory systems.

In this paper, we present the preliminary evaluation of the hypervisor-based
virtualization for the first byte-addressable NVM.
We consider that
RAMinate is the key technology to address a performance gap between DRAM and NVM in the main memory of a computer.
We first clarify the performance characteristics of DCPMM by using micro-benchmark programs.
Next, we applied RAMinate to the hybrid memory system composed of DRAM and DCPMM, and evaluated its feasibility through a preliminary workload.

Section 2 briefly explains the overview of DCPMM and RAMinate.
It also discusses the comparison between DCPMM's Memory Mode and RAMinate.
Section 3 presents the performance characteristics of DCPMM and
the preliminary evaluation of RAMinate for DCPMM.
Section 4 concludes the paper.

\section{Background}
\begin{table}[t]
	\begin{center}
	\caption{The overview of the test machine used in experiments}
	\label{tbl:machine}
	{\footnotesize
	\begin{tabular}{l l}
		\hline
		CPU & Intel Xeon Platinum 8260L 2.40 GHz (Cascade Lake), 2 processors \\
		    & L1d cache 32 KB, L1i cache 32 KB, L2 cache 1024K, L3 cache 36 MB \\
		DRAM & DDR4 DRAM 16 GB, 2666 MT/s, 12 slots \\
		DCPMM & DDR-T 128 GB, 2666 MT/s, 12 slots \\
		OS & Linux Kernel 4.19.16 (extended for RAMinate) \\
		\hline
	\end{tabular}
	}
		\end{center}
\end{table}

\begin{figure}[t]
	\begin{center}
	\includegraphics[width=0.6\textwidth]{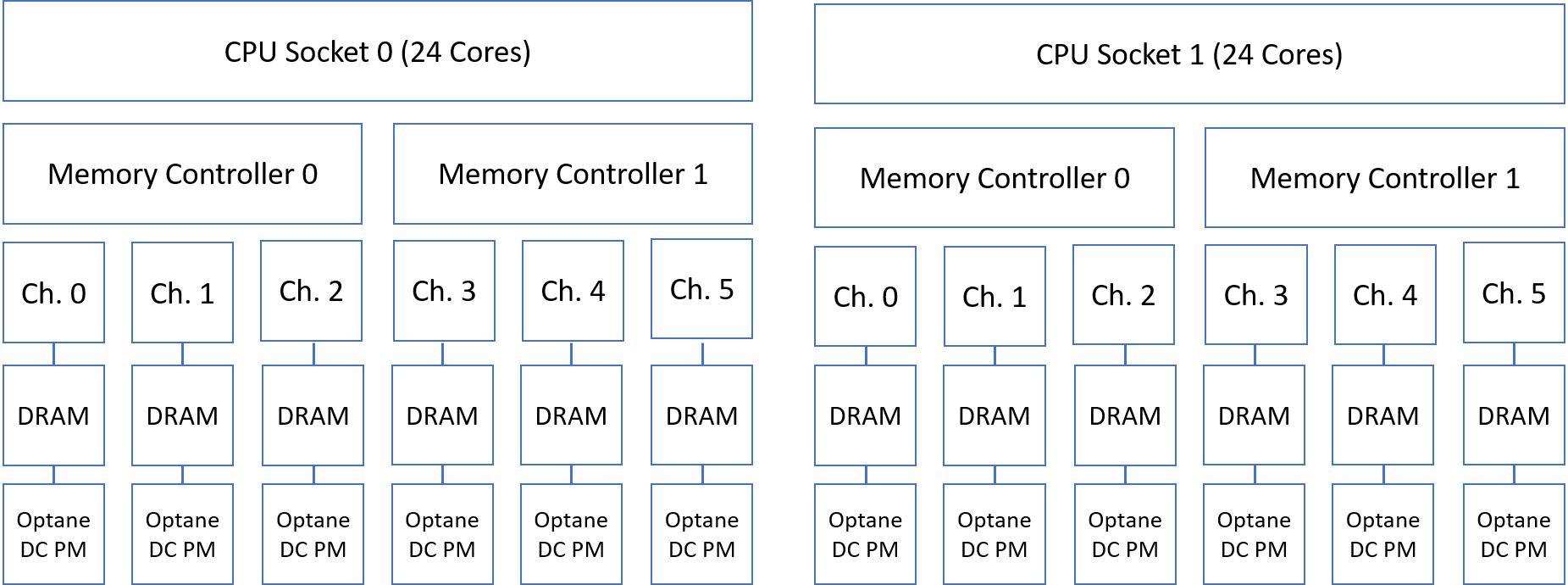}
	\caption{The memory configuration of the tested machine}
	\label{fig:memtopo}
	\end{center}
\end{figure}

\subsection{Intel Optane DCPMM}
Intel Optane DCPMM is an NVM module
connected to the DIMM interface of a computer. The capacity of a DCPMM
is larger than that of ordinary DRAM modules; each DCPMM module has 128 GB in our tested machine.
Its memory cell is based on the 3D XPoint technology, which is a stacked array of resistive memory elements.
Although the details are not disclosed, wear-leveling and buffering are supposed to be performed at the inside of the memory module.
In this report, we focus on the increased capacity of main memory by DCPMM rather than its data persistency.

Table \ref{tbl:machine} summarizes the specification of the tested machine.
Figure \ref{fig:memtopo} shows its memory configuration.
The machine is equipped with 2 CPU sockets. A CPU processor has 24 physical CPU cores and 2 memory controllers.
A memory controller has 3 memory channels. Each memory channel has a DDR4 DRAM module (16 GB) and a DCPMM (128 GB).
The total DRAM size of the machine is 192 GB. The total DCPMM size is 1536 GB.

The Intel CPU processors supporting DCPMM allow users to configure how DCPMM is
incorporated into the main memory of a computer.
In {\bf Memory Mode}, DRAM works as a cache for DCPMMs.
From the viewpoint of the operating system running on the machine,
the total size of the main memory is the sum of DCPMMs.
The memory controller caches read/written data in DRAM, thus reducing performance overhead caused by slow memory access to DCPMM.
This mechanism is implemented in the hardware layer. No modification is necessary to the software layer.
In {\bf App Direct Mode}, the memory controller maps both DRAM and DCPMM to the
physical memory address space of the machine.
It is the responsibility of the software layer to manage the memory space of DCPMM.
Typically, programming libraries and persistent memory file systems are used to take advantage of DCPMM.
Our hypervisor-based virtualization mechanism for hybrid memory systems also works for App Direct Mode.

\subsection{RAMinate}
\begin{figure}[t]
	\begin{center}
	\includegraphics[width=0.5\textwidth]{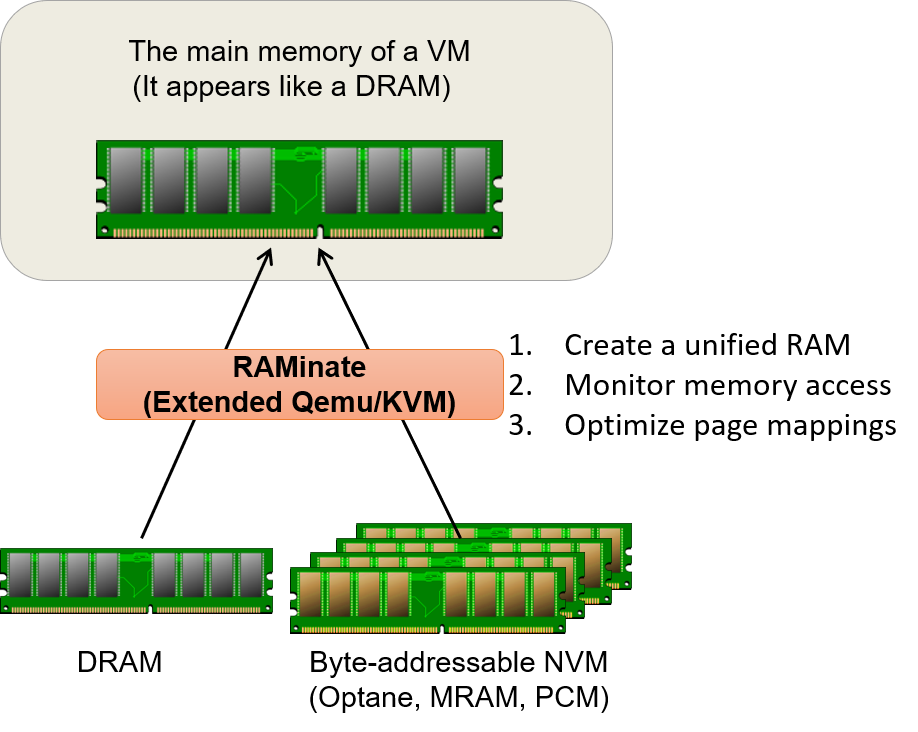}
	\caption{The overview of RAMinate}
	\label{fig:raminate}
	\end{center}
\end{figure}

RAMinate~\cite{raminateurl} is a hypervisor-based virtualization mechanism for hybrid main memory composed of DRAM and byte-addressable NVM.
To our knowledge, RAMinate is the first work implementing hypervisor-based virtualization. In contrast to past studies, our mechanism works at the hypervisor, not at the hardware or operating system level. It does not require any special program at the operating system level nor any design changes of the current memory controller at the hardware level.

RAMinate was originally presented in the seventh ACM Symposium on Cloud Computing 2016
(ACM SoCC 2016). Our paper~\cite{socc2016raminate} obtained the best paper award in the symposium.
In the paper,
we assumed that STT-MRAM (Spin Transfer Torque Magnetoresistive RAM) will be used as a part of main memory.
Since at the moment of ACM SoCC 2016 there was no byte-addressable NVM modules in the market, we evaluated its feasibility by dedicating a part of DRAM to a pseudo STT-MRAM region.
The design of RAMinate is basically independent of the type of an NVM device.
It works for any byte-addressable NVMs (including DCPMM) without any modification to it.

Figure \ref{fig:raminate} illustrates the overview of RAMinate.
It creates a VM and allocates the main memory of the VM from DRAM and
byte-addressable NVM.
From the viewpoint of the guest operating system, the guest main memory appears like
normal memory. Underneath the hood, each memory page of the guest main memory is mapped to DRAM or NVM.
RAMinate periodically monitors memory access of the VM and determines which memory pages are being intensively accessed.
It dynamically optimizes memory mappings in order to place hot memory pages to faster memory (i.e., DRAM in the current system).
RAMinate maximizes the performance of a VM even with a small amount of DRAM mapped to its main memory.
For design details, please refer to our paper~\cite{socc2016raminate}.

\begin{table}[t]
	\begin{center}
	\caption{Comparison between DCPMM's Memory Mode and RAMinate}
	\label{tbl:comparison}
	{\footnotesize
	\begin{tabular}{p{0.22\textwidth}|p{0.34\textwidth}|p{0.34\textwidth}}
		\hline
		\hline
		 & DCPMM's Memory Mode & RAMinate \\
		\hline
		Where the mechanism is implemented & Hardware (i.e., memory controller) & Software (i.e., hypervisor) \\ \hline
		How virtual memory space is created & System wide. The main memory of a physical machine is extended to the size of the DCPMM space assigned to Memory Mode. & Per virtual machine. The main memory of each virtual machine is created by using DRAM and NVM. \\ \hline
		Flexibility in the mixed ratio of DRAM and NVM & The mixed ratio is system-wide configuration. Once changed, reboot is necessary. & Any mixed ratio is possible for each VM. It is possible to dynamically change the mixed ratio without rebooting the VM. \\
		\hline
	\end{tabular}
	}
	\end{center}
\end{table}

Table \ref{tbl:comparison} summarizes the comparison between Memory Mode of DCPMM and RAMinate.
Both the mechanisms provide virtualization for hybrid memory systems.
Since DCPMM's Memory Mode is implemented in a memory controller,
the main memory of a physical machine is extended to the size of the DCPMM space assigned to Memory Mode.
Although it is possible to change the percentage of the DCPMM space to be assigned to Memory Mode,
this configuration is system-wide. Rebooting the physical machine is necessary to apply a new configuration.
On the other hand, RAMinate, implemented in the hypervisor layer, provides greater flexibility.
For each VM, it is possible to specify any mixed ratio of DRAM and NVM.
For example, we can assign a large ratio of DRAM to a VM running a
memory-intensive database server; we can reduce the DRAM ratio of a VM running
batch jobs that do not have tight requirement on service latencies.
An IaaS provider (e.g., Amazon EC2) can increase the density of server consolidation without sacrificing performance of workloads.
As shown in the below,
even though a small amount of DRAM is assigned to a VM,
RAMinate is capable of drastically alleviating performance degradation by dynamically optimizing memory mappings.

\section{Evaluation}
\label{sec:eval}

We first conducted preliminary experiments to measure the basic performance of DCPMM.
The performance of DCPMM is supposed to be different from that of DRAM. We made it clear through our micro-benchmark programs.
Next, we set up RAMinate for DCPMM to see its feasibility for the first byte-addressable NVM.

We assigned all the DCPMMs to App Direct Mode. The host operating system leaves DCPMMs intact.
The benchmark programs directly accessed the physical memory ranges of DCPMMs via the
device file of Linux ({\tt /dev/mem}).
Although the operating system recognized two NUMA domains (i.e., those of CPU
socket 0 and 1, respectively), we used the CPU cores and memory modules only in
the first NUMA domain.

The interleaving mechanism of DRAM and that of DCPMM were enabled, respectively.
For DCPMM, the interleaving configuration of App Direct Mode was used unless otherwise noted.
The 6 DCPMMs connected to each NUMA domain were logically combined.
The memory controller spread memory accesses evenly to the memory modules.
For DRAM, the controller interleaving (i.e., iMC interleaving) was enabled in the BIOS setting.
Similarly, the 6 DRAM modules connected to each NUMA domain were logically combined.
In order to simplify system behavior, we disabled the hyper-threading mechanism of CPUs.
Transparent huge page and address randomization were also disabled in the setting of Linux Kernel.

\subsection{Basic Performance of DCPMM}
We developed micro-benchmark programs that measure the read/write access
latencies and bandwidth of physical memory.
To measure read performance, the micro-benchmark programs induce Last Level Cache (LLC) misses that result in data fetches from memory modules.
For write performance, the programs cause the evictions of modified cachelines as well.

\begin{figure}[t]
	\begin{center}
	\includegraphics[width=0.9\textwidth]{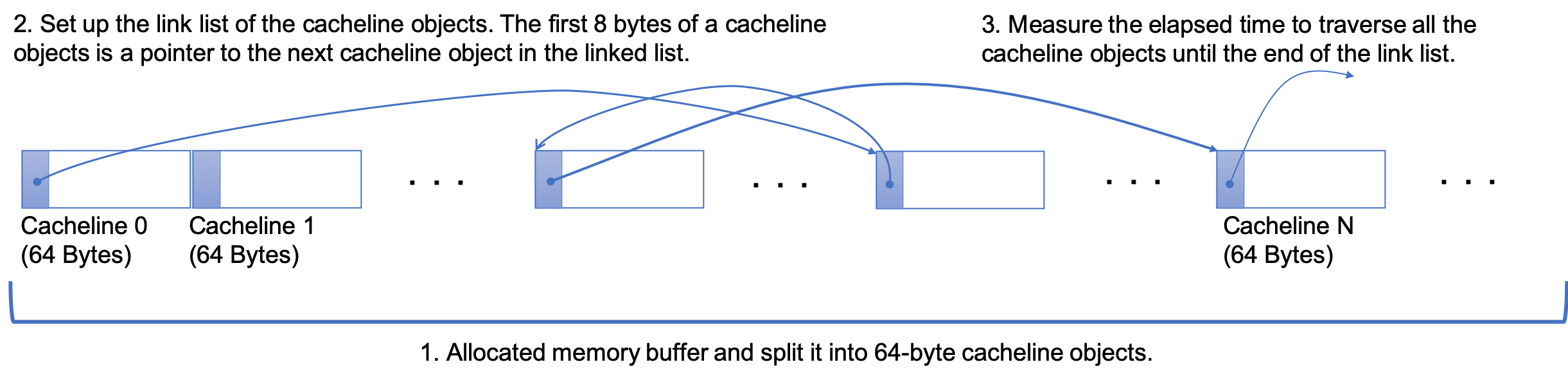}
	\caption{The overview of the micro-benchmark program to measure memory read/write latencies}
	\label{fig:wbbench}
	\end{center}
\end{figure}

\subsubsection{Read/Write Latencies}
\label{sec:explat}

Figure \ref{fig:wbbench} illustrates the overview of the micro-benchmark
program to measure memory read/write latencies.
Most CPU architectures perform the memory prefetching and the out-of-order execution to
hide memory latencies from programs running on CPU cores.
To measure latencies precisely, the benchmark program was carefully designed to suppress these effects.
To measure the read latency of main memory, it works as follows:
\begin{itemize}
	\item First, it allocates a certain amount of memory buffer from a target memory device. To induce LLC misses, the size of the allocated buffer should be sufficiently larger than the size of LLC. It splits the memory buffer into 64-bytes cacheline objects.
	\item Second, it set up the link list of the cacheline objects in a random order, i.e., traversing the linked list causes jumps to remote cacheline objects.
	\item Third, it measures the elapsed time for traversing all cacheline objects and calculates the average latency to fetch a cacheline.
		In most cases, a CPU core stalls due to an LLC miss upon the traversal of the next cacheline object in the linked list. The elapsed time of this CPU stall is a memory latency.
\end{itemize}

When measuring the write-back latency, in addition to the second step, it updates the second 8 bytes of a cacheline object before jumping to the next cacheline object.
The status of the cacheline in LLC changes to {\it modified}. The cacheline is written back to main memory later.
Although a write-back operation is asynchronously performed,
we can estimate the average latency of a memory access involving the write-back of a cacheline, from the elapsed time to traverse all the cache link objects.

\begin{figure}[t]
	\begin{subfigure}{0.5\linewidth}
		\includegraphics[width=\textwidth]{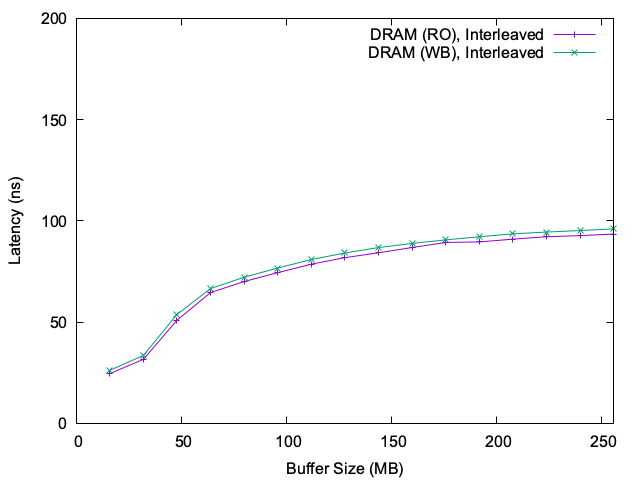}
		\caption{DRAM (Interleaved)}
	\end{subfigure}
	\begin{subfigure}{0.5\linewidth}
		\includegraphics[width=\textwidth]{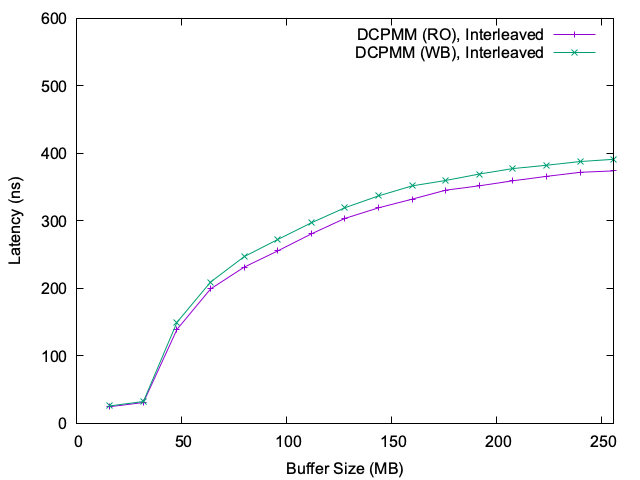}
		\caption{DCPMM (Interleaved)}
	\end{subfigure}
	\begin{subfigure}{0.5\linewidth}
		\includegraphics[width=\textwidth]{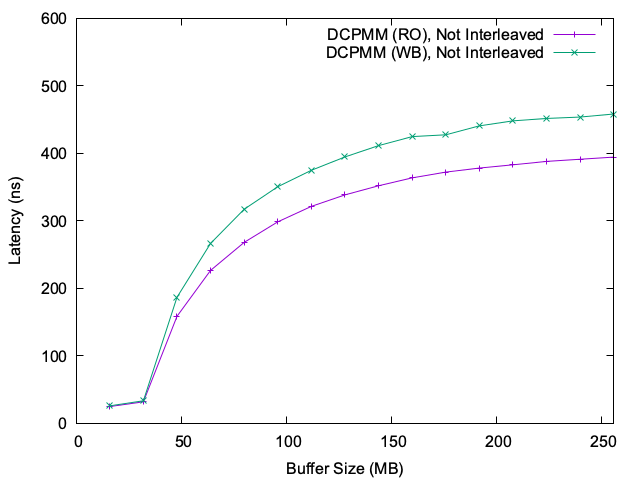}
		\caption{DCPMM (Non Interleaved)}
	\end{subfigure}
	\caption{The read and write latencies of DRAM and DCPMM.
	In the graphs, the results of the read latency are marked as RO
	(read-only), and those of the write latency are marked as WB
	(write-back).}
	\label{fig:memlat}
\end{figure}

Figure \ref{fig:memlat} summarizes the measured results of the read/write latencies of DRAM and DCPMM, respectively.
As the size of the allocated memory buffer increased,
the read/write latencies of DRAM reached approximately 95 ns, respectively.
Although write latencies were slightly higher with any tested buffer sizes,
the differences in read/write latencies were only 1-2 ns.
On the other hand, the read latency of DCPMM was up to 374.1 ns. The write latency was 391.2 ns.
For read access, the latency of DCPMM was 400.1\% higher than that of DRAM. For write access, it was 407.1\% higher.
Similarly to other NVM technologies, the write latency of a bare DCPMM module was larger than the read latency, as clearly shown in the result of the non-interleaved configuration. The latency of memory access involving write-back was 458.4 ns, which was 16.1\% higher than that of read-only access (394.5 ns).

It should be noted that these measured latencies include the penalty caused by
TLB (Translation Lookaside Buffer) misses. The page size in the experiments was 4 KB.
Our measured latencies of DRAM were slightly higher than the value that Intel Memory Latency Checker (MLC) reported.
Intel MLC v3.6 reported that the DRAM latency was 82 ns.
The method of random access in Intel MLC slightly differs from that of our
micro-benchmark program. According to the documentation of Intel MLC v3.6, it performs
random access in a 256-KB range of memory in order to mitigate TLB misses. After
completing that range, it performs random access in the next 256-KB range of
memory.
We consider that memory intensive applications randomly accessing a wide range
of memory will experience memory latencies close to our obtained results.
Although it is out of the scope of this report,
one could use a large page size such as 2 MB and 1 GB to mitigate TLB misses.

\subsubsection{Read/Write Bandwidths}
Our micro-benchmark program measuring the read/write bandwidths of main memory
launches a multiple number of concurrent worker processes to perform memory access.
Each worker process allocates 1 GB of memory buffer from a target memory device.
The memory buffer of a worker process does not overlap the memory buffer of another worker process.
Each worker process sequentially scans its allocated buffer.
We increased the number of worker processes up to the number of CPU cores of an NUMA domain.

\begin{figure}[t]
	\begin{subfigure}{0.5\linewidth}
		\includegraphics[width=\textwidth]{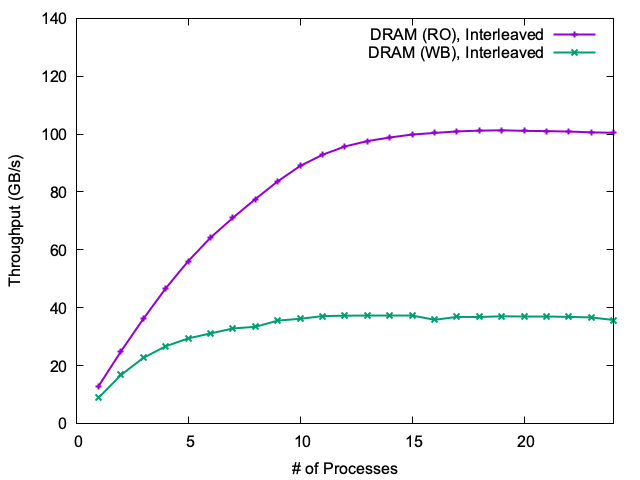}
		\caption{DRAM (Interleaved)}
	\end{subfigure}
	\begin{subfigure}{0.5\linewidth}
		\includegraphics[width=\textwidth]{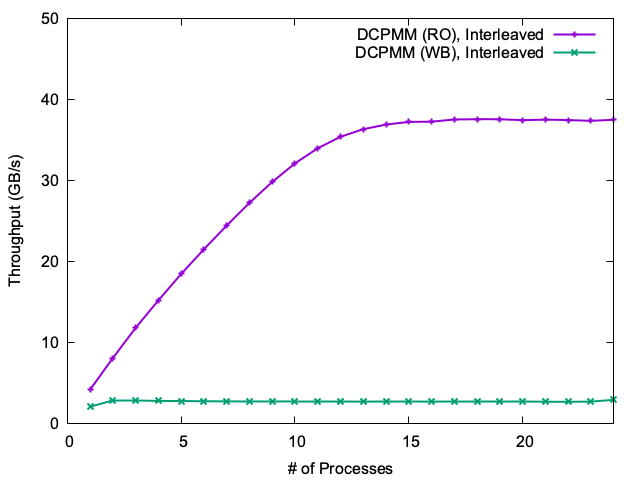}
		\caption{DCPMM (Interleaved)}
	\end{subfigure}
	\begin{subfigure}{0.5\linewidth}
		\includegraphics[width=\textwidth]{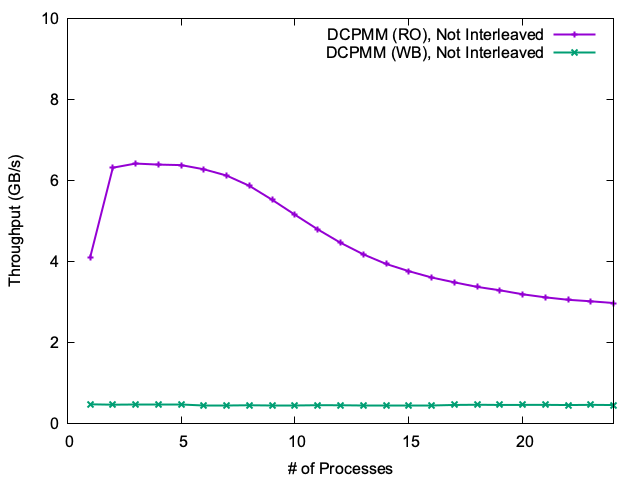}
		\caption{DCPMM (Non Interleaved)}
	\end{subfigure}
	\caption{The read/write memory bandwidths of DRAM and DCPMM. In the graphs, the results of the read latency are marked as RO (read-only), and those of the write latency are marked as WB (write-back).}
	\label{fig:bw}
\end{figure}

Figure \ref{fig:bw} shows the read/write bandwidths of DRAM and DCPMM, respectively.
As the number of the concurrent worker processes increased for read-only memory
access, the bandwidth of DRAM reached 101.3 GB/s at peak; on the other hand, the
bandwidth of DCPMM was 37.6 GB/s.
For memory access involving write-back,
the bandwidth of DRAM was 37.4 GB/s at peak, and that of a DCPMM was 2.9 GB/s.
With the interleaving of DCPMM disabled, the observed peak bandwidths were degraded to approximately 1/6 (i.e., 6.4 GB/s for read-only access, and 0.46 GB/s for write-back-involving access).
The number of the memory modules, being simultaneously accessed, was only one (i.e., 1/6 of the interleaved configuration).

For read access, the throughput of a DCPMM was 37.1\% of DRAM. For write access, it was 7.8\%.
The difference in read and write bandwidths is larger in DCPMM; it was approximately 13 times in DCPMM, while it was 2.7 times in DRAM.

\subsubsection{Summary}
Table \ref{tbl:summary} summarizes the results of our experiments investigating the
performance characteristics of DCPMM.
Although DCPMM provides a large capacity of main memory,
memory access to a DCPMM region is slower than that of DRAM.

Latency:
\begin{itemize}
	\item The read latency was approximately 374.1 ns, which was 400.1\% larger than that of DRAM.
	\item The memory access latency involving write back operations was approximately 391.2 ns, which was 407.1\% larger than that of DRAM. As observed in Section \ref{sec:explat}, without interleaving, it was degraded to approximately 458.4 ns.
\end{itemize}

Bandwidth:
\begin{itemize}
	\item The read bandwidth of DCPMM was approximately 37.6 GB/s, which was 37.1\% of that of DRAM.
        \item The memory access bandwidth involving write back operations was approximately 2.9 GB/s, which was 7.8\% of that of DRAM.
\end{itemize}

\begin{table}[t]
	\begin{center}
		\caption{The summary of the performance characteristics of interleaved DRAM and DCPMM in our experiments}
		\label{tbl:summary}
		{\footnotesize
		\begin{tabular}{l l|r|r|r}
			\hline
			    & & \multicolumn{1}{c}{DRAM} & \multicolumn{1}{c}{DCPMM} & \multicolumn{1}{c}{Ratio} \\ \hline
			\multirow{2}{*}{Latency}              & Read-only  & 93.5 ns   & 374.1 ns   & 400.1\% \\
			                                      & Write-back & 96.1 ns   & 391.2 ns   & 407.1\% \\     \hline
			\multirow{2}{*}{Bandwidth}            & Read-only  & 101.3 GB/s &  37.6 GB/s &  37.1\% \\
			                                      & Write-back &  37.4 GB/s &   2.9 GB/s &   7.8\% \\
			\hline
		\end{tabular}
		}
	\end{center}
\end{table}

\subsection{Preliminary Evaluation of RAMinate with DCPMM}
\begin{figure}[t]
	\begin{center}
	\includegraphics[width=0.9\textwidth]{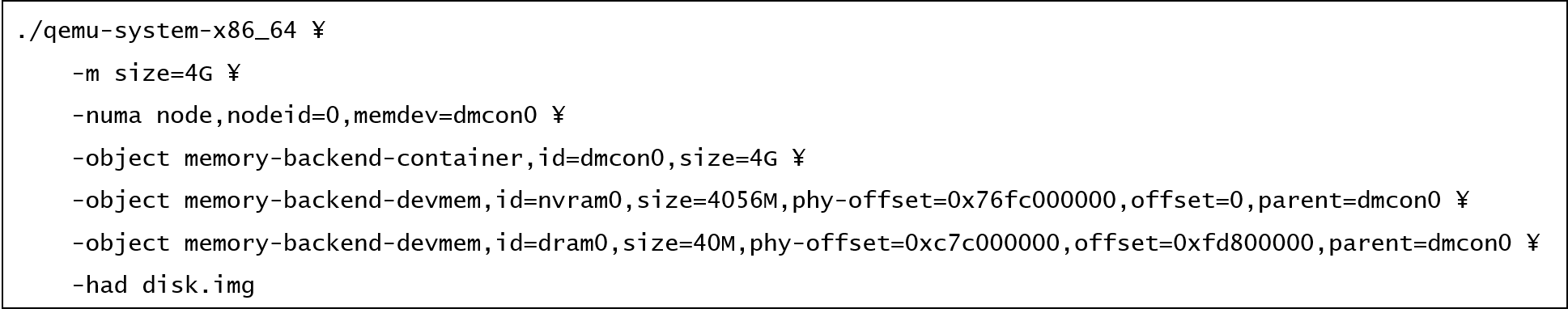}
		\caption{An example command line of RAMinate to create a VM with 4 GB RAM. The mixed ratio of DRAM is 1\% (40 MB).}
	\label{fig:raminatecmd}
	\end{center}
\end{figure}

\begin{figure}[t]
	\begin{subfigure}{0.5\linewidth}
		\includegraphics[width=\textwidth]{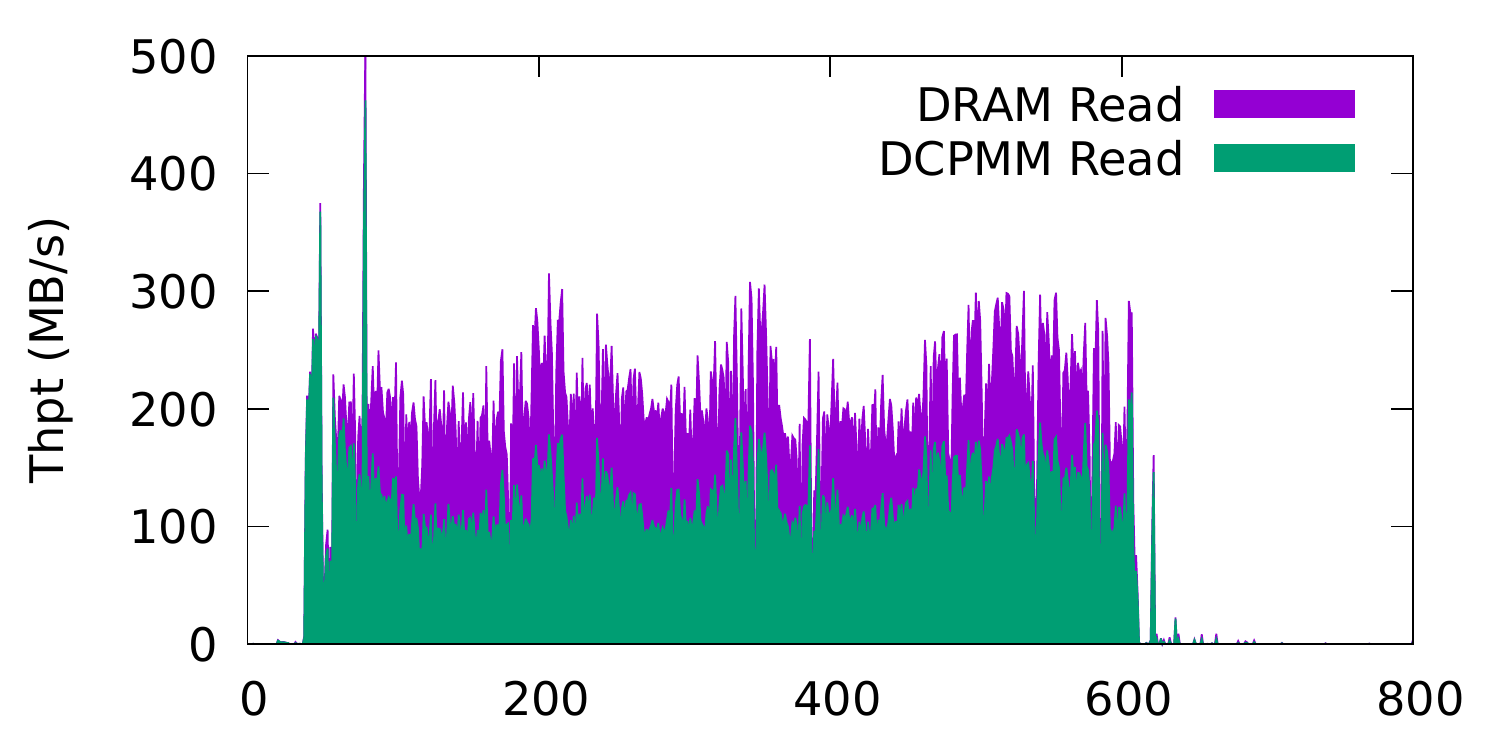}
		\caption{Read}
	\end{subfigure}
	\begin{subfigure}{0.5\linewidth}
		\includegraphics[width=\textwidth]{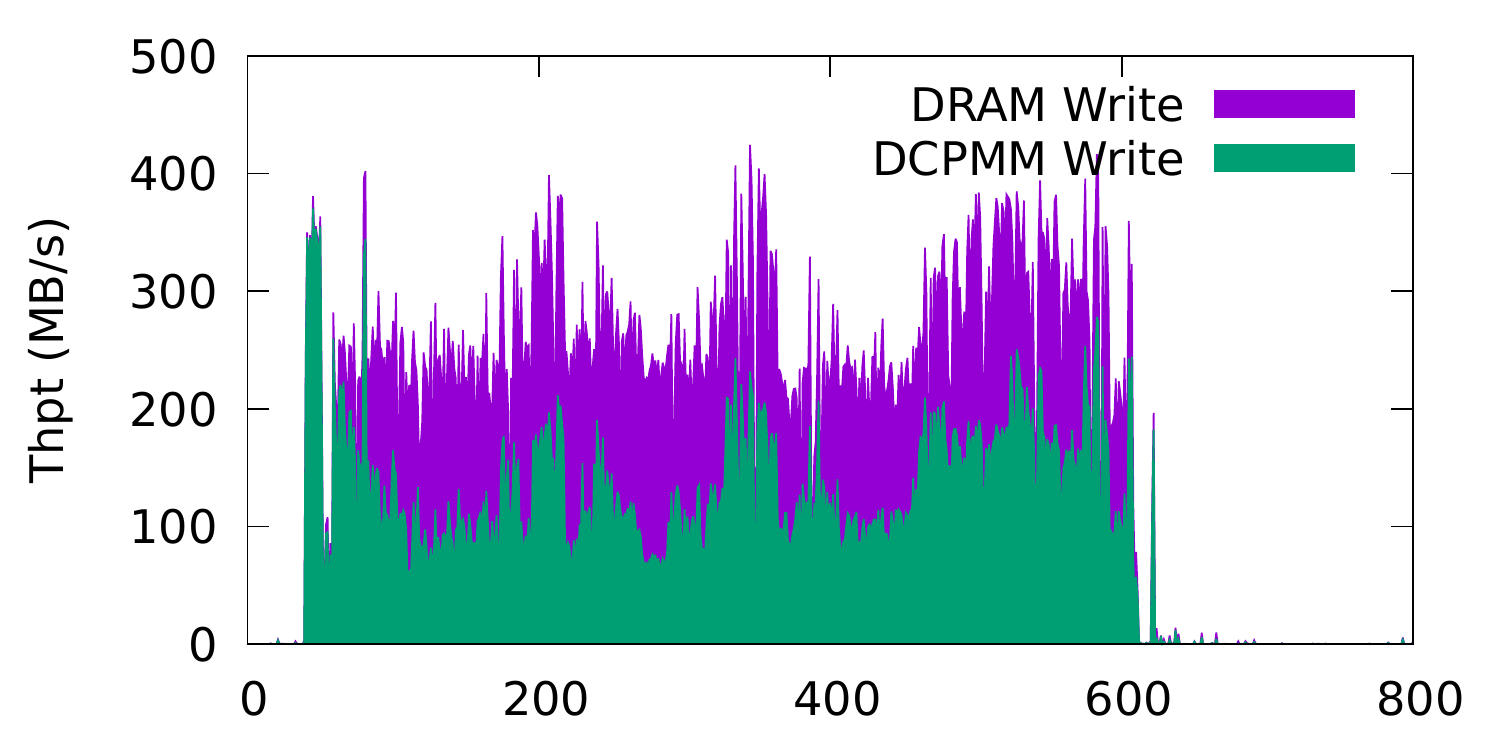}
		\caption{Write}
	\end{subfigure}
	\caption{The read/write traffic of DRAM and DCPMM created by the VM. Kernel build started in the guest operating system at 40 seconds.}
	\label{fig:ramthpt}
\end{figure}

\begin{figure}[t]
	\begin{center}
	\includegraphics[width=0.5\textwidth]{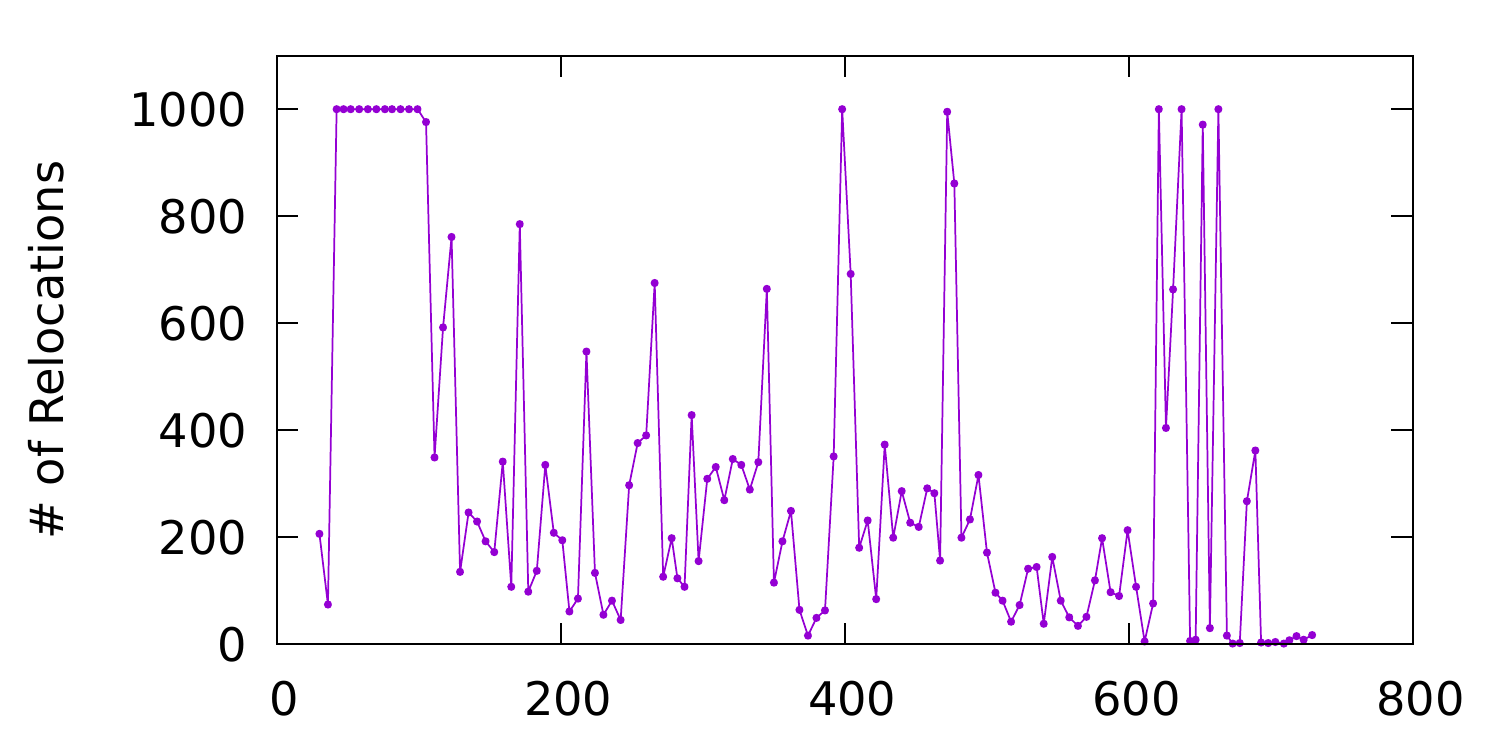}
	\caption{The numbers of page relocations performed by RAMinate during kernel build in the guest operating system.}
	\label{fig:ramnmigs}
	\end{center}
\end{figure}

RAMinate is the key technology to address a performance gap between DRAM and NVM in the main memory of a computer.
As confirmed in the above, there is a substantial performance gap between DRAM and DCPMM.

To confirm its feasibility to DCPMM, we applied RAMinate to the hybrid main memory of the tested machine.
We created a VM with 4 GB RAM composed of 40 MB DRAM and 4056 MB DCPMM (i.e., the mixed ratio of DRAM is 1\%).
To carefully examine how RAMinate spreads memory traffic to both the types of memory devices,
we need to eliminate the impact of memory access of the host operating system.
We therefore reserved the memory modules of Memory Controller 1 of CPU Socket 0 for RAMinate.
We disabled the interleaving of App Direct Mode for DCPMM and assigned a DCPMM region in the DCPMMs of Channel 3-5 to the VM.
We also disabled the interleaving of DRAM memory controllers and assigned a DRAM region in the DRAM modules of Channel 3-5 to the VM.
Figure \ref{fig:raminatecmd} shows the command line of RAMinate to create the VM.
The physical address of {\tt 0x76fc000000} is the start address of a DCPMM region
assigned to the VM, which is mapped to offset {\tt 0} of the guest physical address
of the VM. This DCPMM region is included in the first DCPMM of Memory
Controller 1 of CPU Socket 0.
The physical address of {\tt 0xc7c000000} is the start address of the assigned DRAM region,
which is included in the DRAM modules of the same memory controller.
This DRAM region is mapped to offset {\tt 0xfd800000} (i.e., at 4056 MB) of the guest physical address.

The appropriate percentage of DRAM in the RAM of the VM depends on applications.
In this preliminary report, we set it to a bare minimum value, 1\%, to demonstrate the advantage of memory mapping optimization by RAMinate.
As an example of an application, we built Linux Kernel in the guest operating system of the VM.

During experiments, we monitored the read/write traffic of DRAM and DCPMM, respectively.
The DRAM traffic of the VM was measured by the performance counters of the memory controller.
The DCPMM traffic of the VM was obtained through a utility command of DCPMM (i.e., {\tt ipmctl}).

Figure \ref{fig:ramthpt} shows the read/write traffic of DRAM and DCPMM assigned to the VM.
Just after kernel build started, most memory traffic was generated in the DCPMM region.
However, once RAMinate optimized the locations of hot memory pages, the memory traffic of DCPMM was reduced to 50\%. Please note that the mixed ratio of DRAM is only 1\%.

RAMinate detected hot guest physical pages and moved them to the DRAM region. It also moved cold guest physical pages to the DCPMM region.
In this experiment, RAMinate was configured to swap up to a maximum of 1000 pairs of guest physical pages at once.
The interval of the optimization was set to every 5 seconds.
The numbers of page relocations during the experiment were illustrated in Figure \ref{fig:ramnmigs}.
Initially, RAMinate optimized memory mappings intensively. After 120 seconds,
the numbers of page relocations decreased since most hot memory pages were
already moved to the DRAM region. However, the locations of hot memory pages were ever-changing.
RAMinate continuously updated memory mappings in response to the activity of the workload.

\begin{table}[t]
	\begin{center}
	\caption{The comparison of the elapsed times to finish the build of Linux Kernel. The mixed ratio of DRAM and DCPMM in the 4GB RAM of the VM was changed. The optimization mechanism was enabled/disabled.}
	\label{tbl:keneltime}
	{\footnotesize
	\begin{tabular}{l|l|l}
		\hline
		Mixed Ratio in 4 GB RAM   & Memory Mapping Optimization & Time (s) \\ \hline
		DRAM 100\%                &    -                        &  495     \\ \hline
		DRAM 1\%   and DCPMM 99\% & Enabled                     &  557     \\ \hline
		DRAM 1\%   and DCPMM 99\% & Disabled                    &  624     \\ \hline
		DCPMM 100\%               &    -                        &  633     \\ \hline
	\end{tabular}
	}
		\end{center}
\end{table}

\begin{figure}[t]
	\begin{subfigure}{0.5\linewidth}
		\includegraphics[width=\textwidth]{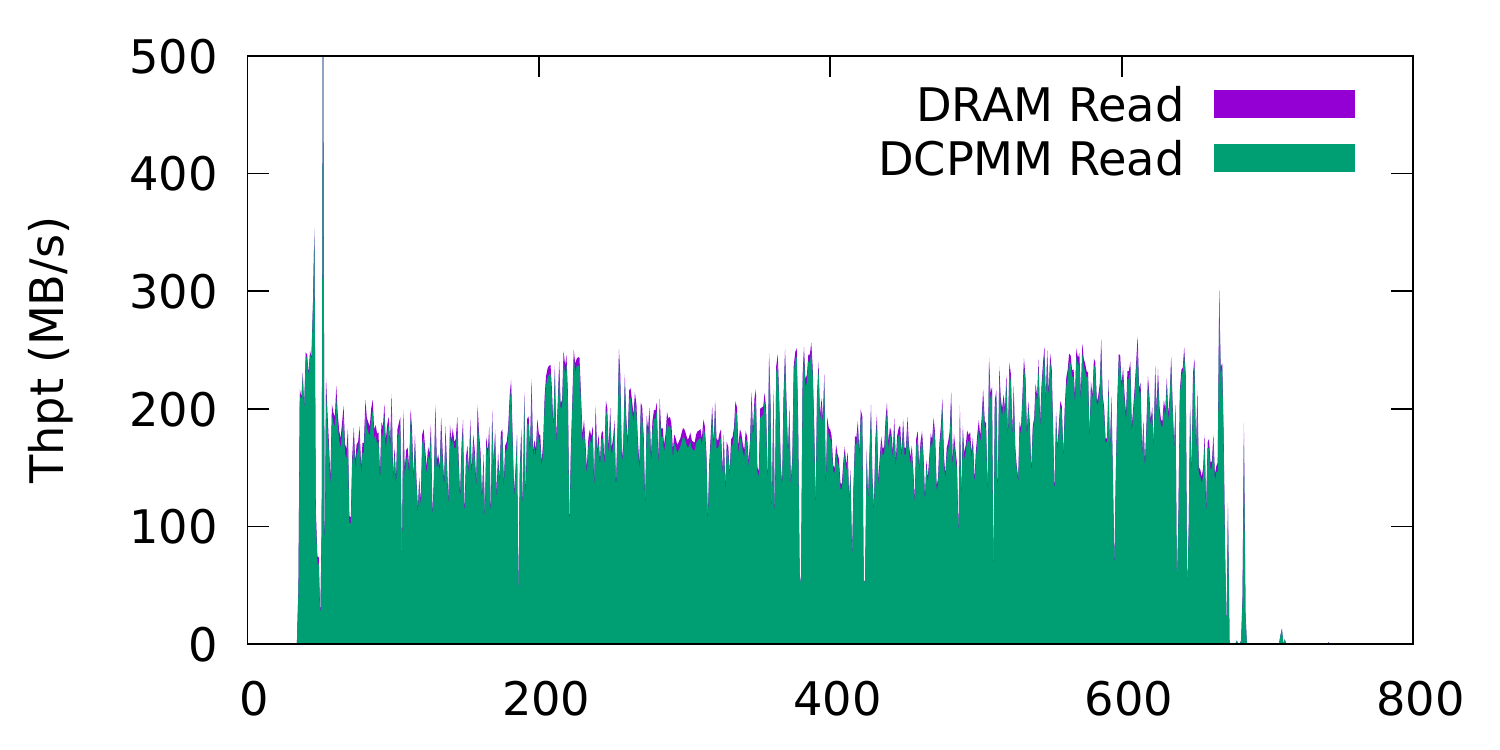}
		\caption{Read}
	\end{subfigure}
	\begin{subfigure}{0.5\linewidth}
		\includegraphics[width=\textwidth]{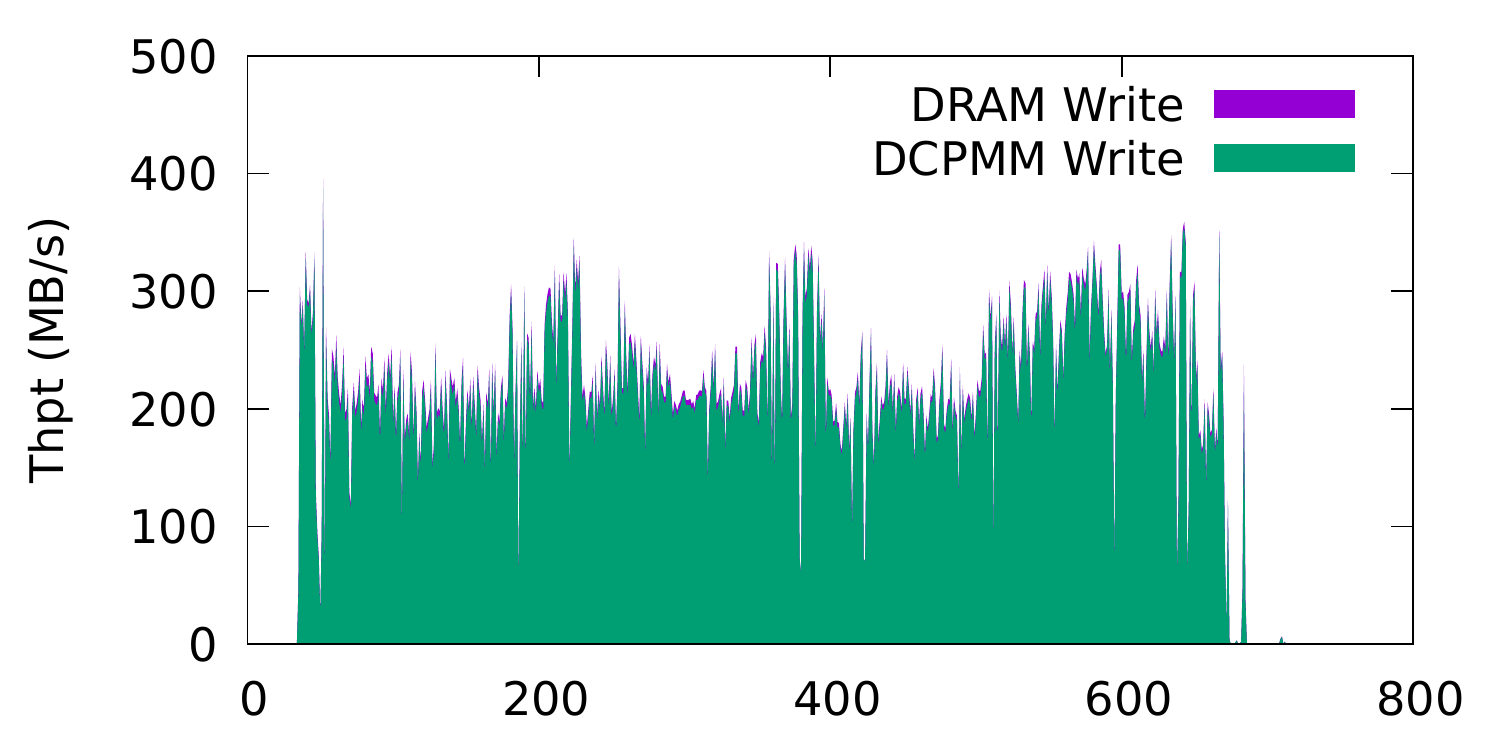}
		\caption{Write}
	\end{subfigure}
	\caption{The read/write traffic of DRAM and DCPMM created by the VM. Kernel build started in the guest operating system at 40 seconds. Since the optimization of memory mappings was disabled during this experiment, the DRAM traffic was merely observed.}
	\label{fig:ramthptnomig}
\end{figure}

The elapsed time to finish the build of Linux Kernel was 557 seconds in the 1\% DRAM case,
which was only 13\% increase from the 100\% DRAM case (i.e., 495 seconds).
When the optimization mechanism was disabled, the elapsed time increased to 624 seconds (i.e. 26\% increase from the 100\% DRAM case).
As shown in Figure \ref{fig:ramthptnomig}, without the optimization, most memory traffic was being generated in
the DCPMM region, thus resulting in serious performance degradation.

In summary, RAMinate successfully created a VM with a specified mixed ratio of DRAM and DCPMM.
Even though the VM had only 1\% of DRAM in its RAM, performance degradation
of the VM was drastically alleviated. RAMinate continuously optimized page
mappings thereby locating hot memory pages to DRAM.

\section{Conclusion}
Intel Optane DCPMM is the first commercially-available, byte-addressable
NVM module connected to the DIMM interface of a computer.
While DCPMM drastically increases the capacity of main memory,
the performance characteristics of DCPMM are substantially different from those of DRAM.
In experiments, we observed that
the read/write latencies of DCPMM were 400\% and 407\% higher than those of DRAM, respectively.
The read/write bandwidths were 37\% and 8\% of those of DRAM.
Therefore,
we believe that our hypervisor-based virtualization mechanism for hybrid main memory systems, RAMinate,
is the key technology to address the performance gap.
In this express report, we confirmed that RAMinate successfully worked for the first byte-addressable NVM.
It improved the performance of a workload by dynamically optimizing memory mappings, placing
hot memory pages to the region of faster memory (i.e., DRAM in the current system).
Even though the VM had only 1\% of DRAM in its RAM, performance degradation
of the VM was drastically alleviated.
The elapsed time to finish the build of Linux Kernel was 557 seconds.
which was only 13\% increase from the 100\% DRAM case (i.e., 495 seconds).
When the optimization mechanism was disabled, the elapsed time increased to 624
seconds (i.e. 26\% increase from the 100\% DRAM case).
We are conducting additional experiments to thoroughly examine the
feasibility of RAMinate under various conditions with DCPMM. Further details will be reported in future publications.

We would like to acknowledge the support of Intel Corporation.
We also thank Dr. Jason Haga and other colleagues for their invaluable feedback.

\bibliographystyle{unsrt}
\bibliography{ms}

\end{document}